\documentstyle[11pt,epsfig]{article}
\begin{document}
\setlength{\baselineskip}{0.30in}
\newcommand{\beq}{\begin{equation}}
\newcommand{\eeq}{\end{equation}}
\newcommand{\bi}{\bibitem}
\newcommand{\cd}{\cdot}

{\hbox to\hsize{August, 1997 \hfill TAC-1997-30}
\begin{center}
\vglue .06in
{\Large \bf {On the possibility of distinguishing between
Majorana and Dirac neutrinos}}
\bigskip
\\{\bf S.H. Hansen\footnote{e-mail: sthansen@tac.dk}
 \\[.05in]
{\it{Teoretisk Astrofysik Center\\
Juliane Maries Vej 30, DK-2100, Copenhagen, Denmark}}} \\[.40in]
\end{center}

\begin{abstract}
We clarify that one cannot distinguish between Majorana and Dirac
neutrinos in the limit of vanishing neutrino mass.
In particular we show, that
the forward-backward asymmetry in the reaction
$e^+ e^- \rightarrow \nu \nu$ is the same for $\nu_M$ and $\nu_D$,
in contrast to recent claims made in the literature.
\end{abstract}

\section{Introduction}
If the neutrinos are massive~\cite{mo}, then they are described by either 
Dirac fields with four independent components or by Majorana fields
with only two, $\nu_L$ and $\hat{\nu}_R = (\nu_L)^C$. 
This possibility of being a Majorana particle arises if the neutrinos 
have no conserved charges so that one cannot distinguish between particles 
and antiparticles.

In the strictly massless case two of the standard model Dirac components are
sterile to the weak interactions and the Dirac field is equivalent 
to the two component Weyl fermions describing the Majorana field.
Thus there is no distinction between $\nu_M$ and $\nu_D$~\cite{st}.

Since the neutrino mass only enters the Lagrangian as $\sum_{a,b} 
\overline{\psi}_{aL} M_{ab} \hat{\psi}_{bR} + h.c.$, one would
consider the mass as being a perturbation to the massless case
and hence expect physical observables to be smooth in the limit
$m_\nu \rightarrow 0$. In this letter we will show that this
is indeed the case for the forward-backward asymmetry, and thereby
clarifying some confusion in the recent literature~\cite{ah}.

\section{The massless limit}
It has long been known, that there is no distinction between Majorana 
and Dirac neutrinos, when the neutrino mass is zero, 
$m_\nu = 0$~\cite{mo, st}.
Various authors have described the difference between $\nu_M$ and $\nu_D$
in the massive case~[4-8], however, there still seem to be some
confusion in the understanding of the $m_\nu \rightarrow 0$ limit.
Indeed, it was recently claimed~\cite{ah} that this limit is not smooth 
and the difference could be seen in the forward-backward asymmetry in
the process $e^+ e^- \rightarrow \nu \nu$.
We will show in the following, that the limit $m_\nu \rightarrow 0$
is smooth.

In order to find the differential cross section for the neutral 
current\footnote{In the charged current processes the limit is also 
smooth~\cite{ga}.}  
process $e^+ e^- \rightarrow \nu_M  \nu_M$ (or $ e^+ e^- \rightarrow \nu_D  
\overline{\nu}_D$) depicted in Fig.~1,
we can evaluate the expressions for the matrix elements in 
eqn.~(\ref{eqn:majorana}) in the appendix for the
Majorana case and eqn.~(\ref{label:dirac}) for Dirac neutrinos.
We go to  the CM-frame~\cite{yt}  using the 
coordinate system of Fig.~\ref{figkoor}, 
\begin{figure}[htb]
\begin{center}
  \begin{minipage}{60mm}
    \centerline{
    \epsfxsize=40mm
    \epsfbox{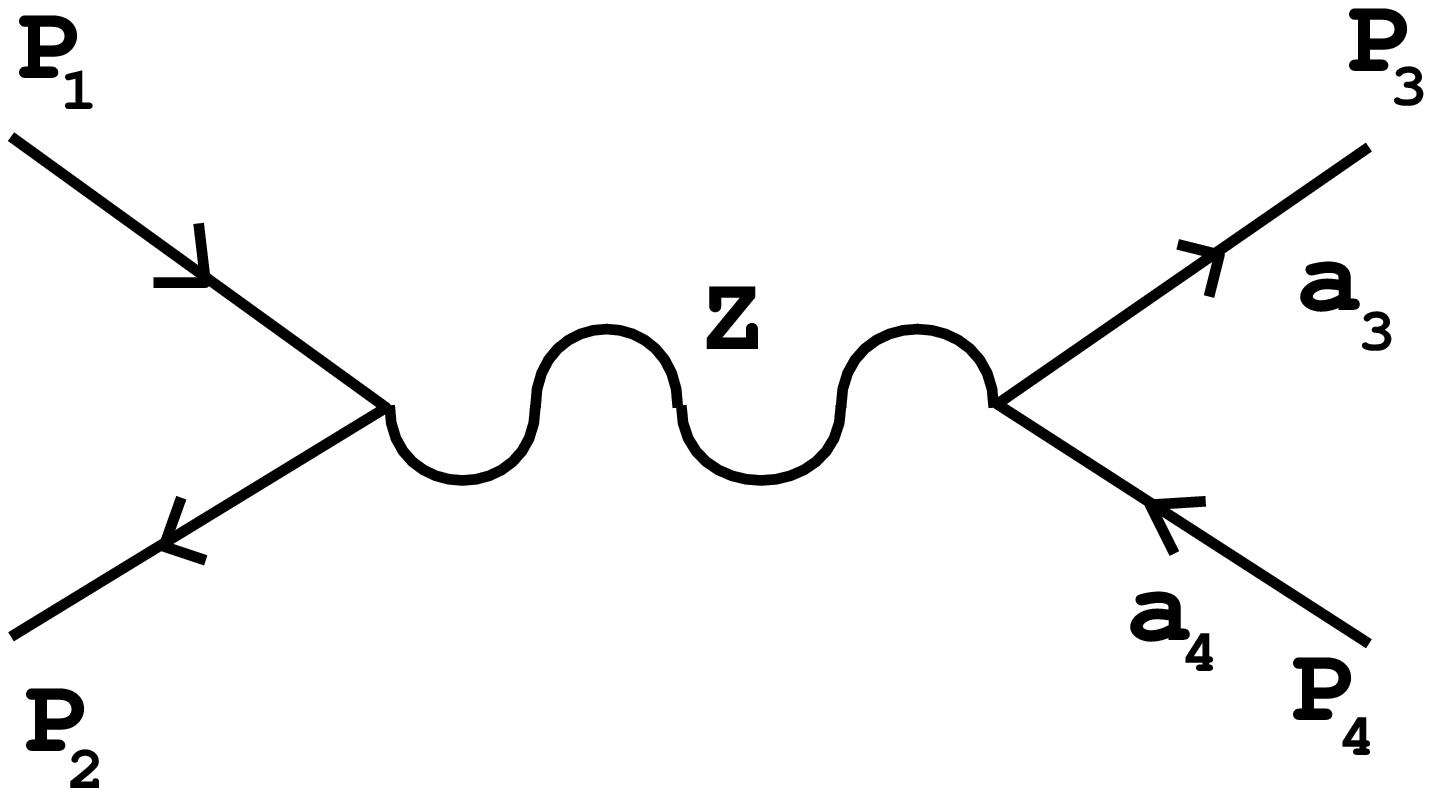}    }
%    \caption{$a$}
  \end{minipage}
  \begin{minipage}{60mm}
    \centerline{
    \epsfxsize=40mm
    \epsfbox{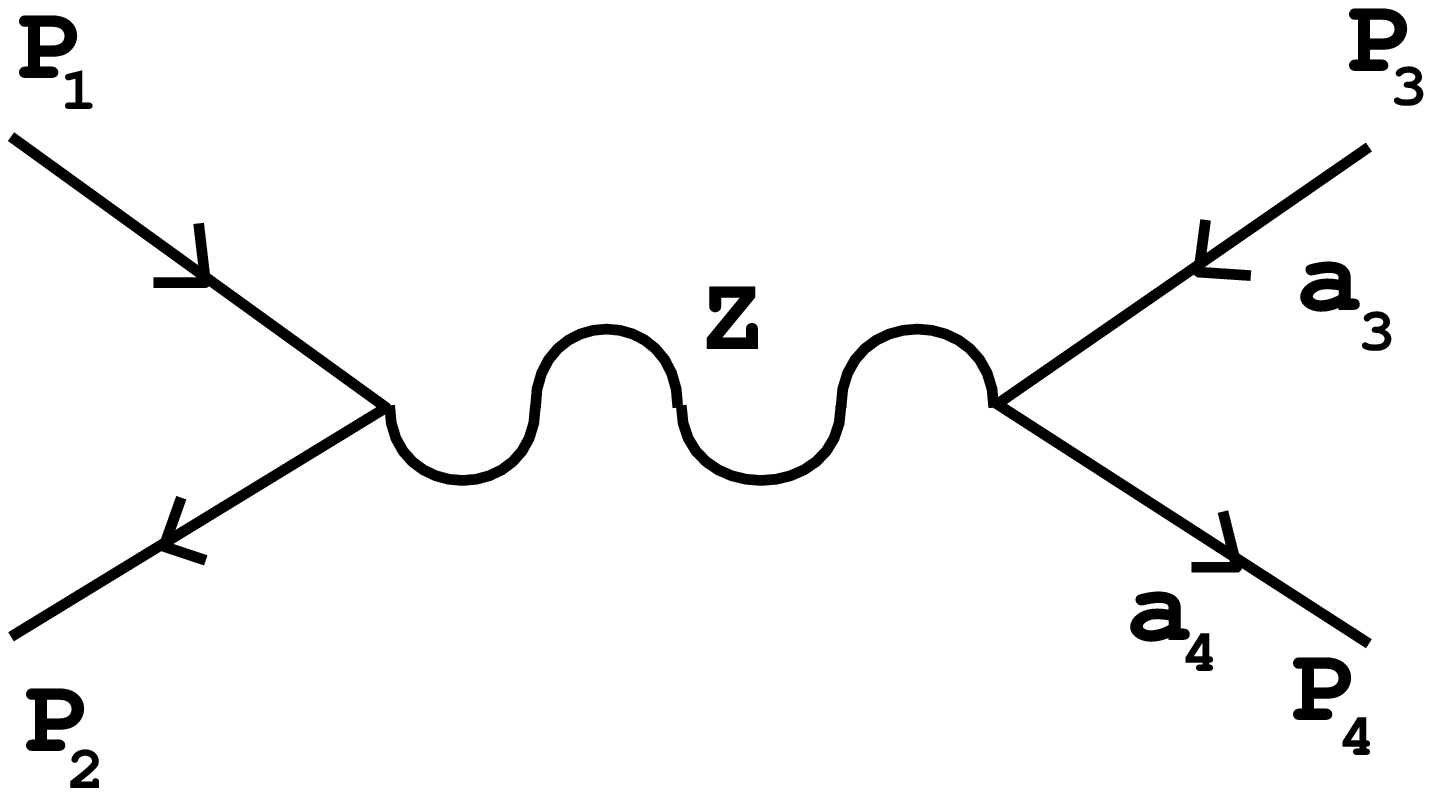}    }
%    \caption{$a = 0.15$}
  \end{minipage}
\caption{Feynmann diagrams and notation for the reaction $e^+ e^- 
\rightarrow \nu_M \nu_M$. For Dirac neutrinos there is only one diagram.}
\end{center}\label{figmaj}
\end{figure}
\begin{figure}[htb]
\begin{center}
\leavevmode
\hbox{
\epsfysize=1.5in
\epsffile{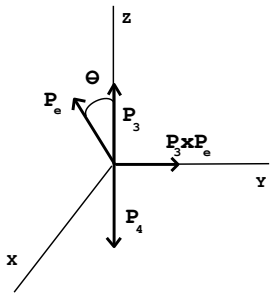}}
\caption{Coordinate system and notation for the calculation of the
differential cross section for the reaction $e^+ e^- \rightarrow \nu \nu$.}
\end{center}\label{figkoor}
\end{figure}
where the left-handed
neutrino always is along the positive z-axis, both for Majorana and
for Dirac neutrinos. With this notation 
the neutrino and electron four-momenta are:
\[
p_{\nu} = p_3 = E(1,0,0,\beta) \; \; \; \;  \mbox{and} \; \; \; \;
p_{e} = p_1 = E_e(1,\sin{\theta}, 0, \cos{\theta})
\]
and the polarization four-vector which reduces to the unit spin vector
in the particle rest frame is:
\[
a_{\nu} =a_3 = (\beta \gamma s_z, s_x, s_y, \gamma s_z),
\]
with $\beta = (1 - (m_\nu/E)^2)^{1/2}$ and $\gamma = E/m_\nu$.
In the limit of $\beta \rightarrow 1$ the Dirac neutrinos have their
spins along the z-axis, and the differential cross
section at the Z-peak smoothly approaches:
\beq
\label{eqn:dir}
\left.  \frac{d \sigma _D}{d \Omega}  \right|_{m \rightarrow 0}
= \sigma_0 \; \left[ \left( g_L^2 + g_R^2 \right) 
\left( 1 + \cos^2{\theta} \right) + 2 \left( g_L^2 - g_R^2 \right) 
\cos{\theta} \right] (1 - s_z)(1 - \overline{s}_z),
\eeq
where $g_L = -1/2 + \sin ^2 \theta_W$, $g_R =  \sin ^2 \theta_W$ and
$\theta$ is the angle between the electron and the left-handed neutrino:
\[
\sigma_0 =  \frac{G_F^2 s}{32 \pi^2} \left| \frac{M_Z^2}{s - M_Z^2 + 
i M_Z\Gamma_Z} \right|.
\]
This is describing left-handed neutrinos (not the anti-neutrinos) 
going in the positive z-direction. 
That is, in a given experiment we would primarily 
measure electrons in the upward direction

In the Majorana case we should be slightly more careful. Let us for
the moment ignore the transverse degrees of freedom, $s_x$ and $s_y$, 
which are unphysical for the massless neutrinos\footnote{A massive 
majorana neutrino is in general not in a helicity eigenstate, however,
the terms including $s_x$ and $s_y$ are proportional to $\cos ^2 \theta$
and do therefore not contribute to the forward-backward asymmetry.
For inclusion of the transverse degrees of freedom see refs.~\cite{yt, mp}.}. 
Then the Majorana differential cross section at the Z-peak
in the $\beta \rightarrow 1$ limit is:
\beq
\label{eqn:maj1}
\left.  \frac{d \sigma _M}{d \Omega}  \right|_{m \rightarrow 0}
= \sigma_0 \; \left[ 
 \left( g_L^2 + g_R^2 \right) 
\left( 1 + \cos^2{\theta} \right) (1 + s_z \overline{s}_z)
+ 2 \left( g_L^2 - g_R^2 \right) 
\cos{\theta} \left( s_z + \overline{s}_z \right)
\right],
\eeq
which can be rewritten as:
\begin{eqnarray}
\label{eqn:maj}
\left.  \frac{d \sigma _M}{d \Omega}  \right|_{m \rightarrow 0}
 = &1/2 \; \sigma_0 \;   \left[ \left( g_L^2 + g_R^2 \right) 
\left( 1 + \cos^2{\theta} \right) +  2 \left( g_L^2 - g_R^2 \right) 
\cos{\theta} \right] \left( 1 - s_z \right) \left(1 -  \overline{s}_z \right) 
\nonumber \\
&+1/2  \; \sigma_0 \;   \left[ \left( g_L^2 + g_R^2 \right) 
\left( 1 + \cos^2{\theta} \right) -  2 \left( g_L^2 - g_R^2 \right) 
\cos{\theta} \right] \left( 1 + s_z \right) \left(1 + \overline{s}_z \right) .
\end{eqnarray}
This expression includes both the positive and the negative helicity
states in the upward direction, in contrast to the Dirac result in
eqn.~(\ref{eqn:dir})
where we only had the negative helicity states. One could say, that
in the Dirac case we only have $\nu$'s upwards, whereas we have 
both $\nu$ and $\overline{\nu}$ in the Majorana case. However, in the
Majorana case these two terms do not cancel, but add up to give the same
cross section as in the Dirac case.

More precisely, the first term of eqn.~(\ref{eqn:maj}), which is exactly 
half of the Dirac 
expression in eqn.~(\ref{eqn:dir}) (as pointed
out in ref.~\cite{mp}), describes the probability
for emitting the left-handed neutrino along the positive z-direction,
and the second term of eqn.~(\ref{eqn:maj}) gives the probability for 
emitting the right-handed
neutrino in the positive z-direction. However, the definition of the
coordinate system in Fig.~\ref{figkoor}, which is for both Majorana and 
Dirac particles, demands that the left-handed neutrino is in the positive 
z-direction, and we must therefore
interchange $z \rightarrow -z$ in the last term of eqn.~(\ref{eqn:maj}). 
This also implies that $\cos{\theta} \rightarrow - \cos{\theta}$, and
therefore the right-handed neutrinos contribute the same to
$\frac{d \sigma _M}{d \Omega}$ as the left-handed.
Thus the differential cross sections for $\nu_M$ and $\nu_D$ 
are the same in the massless limit.
In particular the forward-backward asymmetry~\cite{lr}, 
$A_{FB} \approx 0.45$, is the
same for Dirac and Majorana particles in the reaction $e^+ e^- 
\rightarrow \nu \nu$.

\section{Conclusion}
We have calculated the differential cross sections for the interaction 
$e^+ e^- \rightarrow \nu \nu$ with either Majorana or Dirac neutrinos
in the final state. We showed, that these cross sections are identical
in the $m_\nu \rightarrow 0$ limit. In particular we have explained
why the asymmetric terms don't cancel in the Majorana case, but instead
add up to give the same as in the Dirac case.\\

\bigskip

{\bf Acknowledgment.}
The author is very grateful to professor A. Dolgov for drawing the
attention to this problem and for illuminating conversations.
The work was supported in part by the Danish National Science Research 
Council through its support to the establishment of the Theoretical 
Astrophysical Center. 

\appendix

\section{Matrix elements}
\label{sec:matrixelements}

The differential cross section for the neutral current interaction in
Fig.~1 is proportional to the polarization 
function~\cite{ga}:
\[
\frac{d \sigma}{d \Omega}  \sim G_F^2 \; 
Tr \left( \rho _3 {\cal O}_\alpha \rho_{4} {\cal O} ^\beta
\rho_{2} {\cal O}^\alpha \rho_{1} {\cal O}_\beta \right),
\]
where e.g. ${\cal O}^\alpha = 1/2 \; \gamma^\alpha (1 \pm \gamma_5)$
for Dirac fermions, and because the Majorana field is CPT invariant it has
only axial coupling, ${\cal O}^\alpha = \gamma^\alpha \gamma_5$~\cite{ks}.
One must use the polarization density matrix:
\begin{equation}
\rho_i = 1/2 \; (\not p_i \pm m)(1 - \gamma_5 \not a_i), \nonumber
\end{equation}
when the particles are partially polarized and we measure the spins 
of both the resulting particles in a
reaction such as $e^+ e^- \rightarrow \nu \nu$.
Here we are using the polarization four-vector:
\[
a_\mu = (|\vec{p}|/m \; s_{\parallel}, \vec{s}_{\perp}, E/m \; s_{\parallel} )
= (\beta \gamma  s_{\parallel}, \vec{s}_{\perp}, \gamma  s_{\parallel} ),
\]
which in the particles rest frame reduces to the unit spin vector and obeys 
$(p \cd a) = 0$. 

For the Majorana case the matrix element is:
\begin{eqnarray}
\label{eqn:majorana}
&&|A|^2_M = \\
&&  2^{6} G_F^2 \left[ 2 m_e^2 g_L g_R \left( (p_3 \cd p_4) - m_\nu^2 (a_3 \cd 
a_4) -2 m_\nu^2 \right)  \right. \nonumber \\ 
&&+ m_\nu \; (g_L^2 - g_R^2) \left(p_1 \cd (a_4 - a_3) p_2 \cd (p_4 - p_3)  -
p_2 \cd (a_4 - a_3) p_1 \cd (p_4 - p_3) \right) \nonumber \\
&& +  (g_L^2 + g_R^2) \left[ \left( (p_1 \cd p_3) (p_2 \cd p_4) 
+ (p_1 \cd p_4) (p_2 \cd p_3) \right)
\left( 1 + (a_3 \cd a_4 ) \right) \right. \nonumber \\
&& - m_\nu ^2 \left( (p_1 \cd a_3) (p_2 \cd a_4)
+ (p_1 \cd a_4)(p_2 \cd a_3) \right) \nonumber \\
&&- (p_1 \cd p_2) \left((a_3 \cd a_4)(p_3 \cd p_4) - (a_3 \cd p_4)(p_3 \cd 
a_4) + m_\nu^2  \right) \nonumber \\
&&+ (p_3 \cd p_4) \left((a_3 \cd p_2) (a_4 \cd p_1) + (a_3 \cd p_1)(a_4 \cd 
p_2) \right) \nonumber \\
&& -(p_3 \cd a_4) \left(a_3 \cd p_2)(p_4 
\cd p_1) + (a_3 \cd p_1 ) (p_4 \cd p_2)   \right) \nonumber \\
&& \left. \left.- (a_3 \cd p_4) \left((a_4 \cd p_2)(p_3 \cd p_1) 
+ (a_4 \cd p_1)(p_3 \cd  p_2) \right)
\right] \right], \nonumber
\end{eqnarray}
and for the Dirac case it is:
\begin{eqnarray}
\label{label:dirac}
&&|A|^2_D = \\
&&  2^{6} G_F^2 \left[ 2 m_e^2 g_L g_R \left( (p_3 \cd p_4)
- m_\nu (p_3 \cd a_4) + m_\nu (a_3 \cd p_4) - m_\nu^2 (a_3 \cd a_4) \right) 
\right.  \nonumber \\
&& + (g_L^2 - g_R^2) \left[ (p_1 \cd p_3) (p_2 \cd p_4) - (p_2 \cd p_3)
(p_1 \cd p_4) \right. \nonumber \\
&& - m_\nu^2  \left( (p_1 \cd a_3)(p_2 \cd a_4) - 
(p_2 \cd a_3)(p_1 \cd a_4) \right) \nonumber \\
&& + m_\nu \; \left. \left( (p_1 \cd a_3)(p_2 \cd p_4) - 
(p_2 \cd a_4)(p_1 \cd p_3)
+ (p_2 \cd a_3)(p_1 \cd p_4) - (p_1 \cd a_4)(p_2 \cd p_3) \right) \right]
\nonumber \\
&& + (g_L^2 + g_R^2) \left[ (p_1 \cd p_3) (p_2 \cd p_4) + (p_2 \cd p_3)
(p_1 \cd p_4) \right. \nonumber \\
&&- m_\nu^2 \left((p_1 \cd a_3)(p_2 \cd a_4) + (p_2 \cd a_3)(p_1 \cd a_4)
  \right) \nonumber \\
&& + \left. \left. m_\nu \; \left( (p_1 \cd a_3)(p_2 \cd p_4) 
- (p_2 \cd a_4)(p_1 \cd  p_3)
- (p_2 \cd a_3)(p_1 \cd p_4) + (p_1 \cd a_4)(p_2 \cd 
p_3) \right) \right] \right] . \nonumber
\end{eqnarray}

\end{document}